\documentclass{appolb}
\newcommand{\AGeV}[1][ ]{$A$~GeV{#1}}

\newcommand{\T}{$T$}

\usepackage{epsfig}

\begin{document}
\title{Strangeness Excitation Functions and Transition from Baryonic to Mesonic Freeze-Out.}%
\thanks{Presented at EMMI Workshop and XXVI Max Born Symposium, University of Wroclaw, Poland, July 9 - 11,2009}%
\author{Jean~Cleymans,
\address{UCT-CERN Research Centre and Department of Physics, University of Cape Town, Rondebosch 7701, South Africa,}
\and
Helmut Oeschler,
\address{Darmstadt University of Technology, D-64289 Darmstadt, Germany,}
\and
Krzysztof Redlich,
\address{Institute of Theoretical Physics, University of Wroc\l aw,
Pl-45204 Wroc\l aw, Poland.\\
GSI Hemholtzzentrum f¨ur Schwerionenforschung, D-64291 Darmstadt, Germany.}
\and Spencer Wheaton
\address{UCT-CERN Research Centre and Department of Physics, University of Cape Town, Rondebosch 7701, South Africa.}
}
\maketitle
\begin{abstract}
The sharp peak in the $K^+/\pi^+$ ratio
in relativistic heavy-ion collisions is
discussed in the framework of the Statistical Model. In this model
a rapid change is expected as the hadronic gas undergoes a
transition from a baryon-dominated  to a meson-dominated gas.
The maximum in the
$\Lambda/\pi$ ratio is well reproduced by the Statistical Model,
but the change in the $K^+/\pi^+$ ratio is somewhat less pronounced
than the one observed by the NA49 collaboration. The
calculated smooth increase of the $K^-/\pi^-$ ratio and the shape
of the $\Xi^-/\pi^+$ and $\Omega^-/\pi^+$ ratios
exhibiting maxima at different incident energies is
consistent with the presently available experimental data. We
conclude that the measured particle ratios with $20-30\%$ deviations
agree with a hadronic freeze-out scenario. These deviations seem
to occur just in the transition from baryon-dominated to
meson-dominated freeze-out.
\end{abstract}
\PACS{12.40.Ee, 25.75.Dw}
\section{Introduction}
The NA49 Collaboration  has obtained
yields of strange particles in central Pb-Pb collisions at
20, 30, 40, 80 and 158 $A$ GeV beam energies
\cite{NA49,Gazdzicki,Lambda-NA49,:2007fe,:2008vb}. An
interesting result from this energy scan is the pronounced maximum in
the $K^+/\pi^+$
and $\Lambda/\pi$ ratios around 30 \AGeV beam energy.
Such a behaviour, which is not observed in $p-p$ collisions, has
lend support to speculations that a phase transition
has been seen in heavy ion collisions~\cite{Gorenstein,Stock}.
A more conventional interpretation has
been presented within the hadron gas model~\cite{max_strange} which
describes the results qualitatively very well with the possible exception of the
sharpness of the peak.
%
\begin{figure}[htb]
\centerline{\epsfig{file=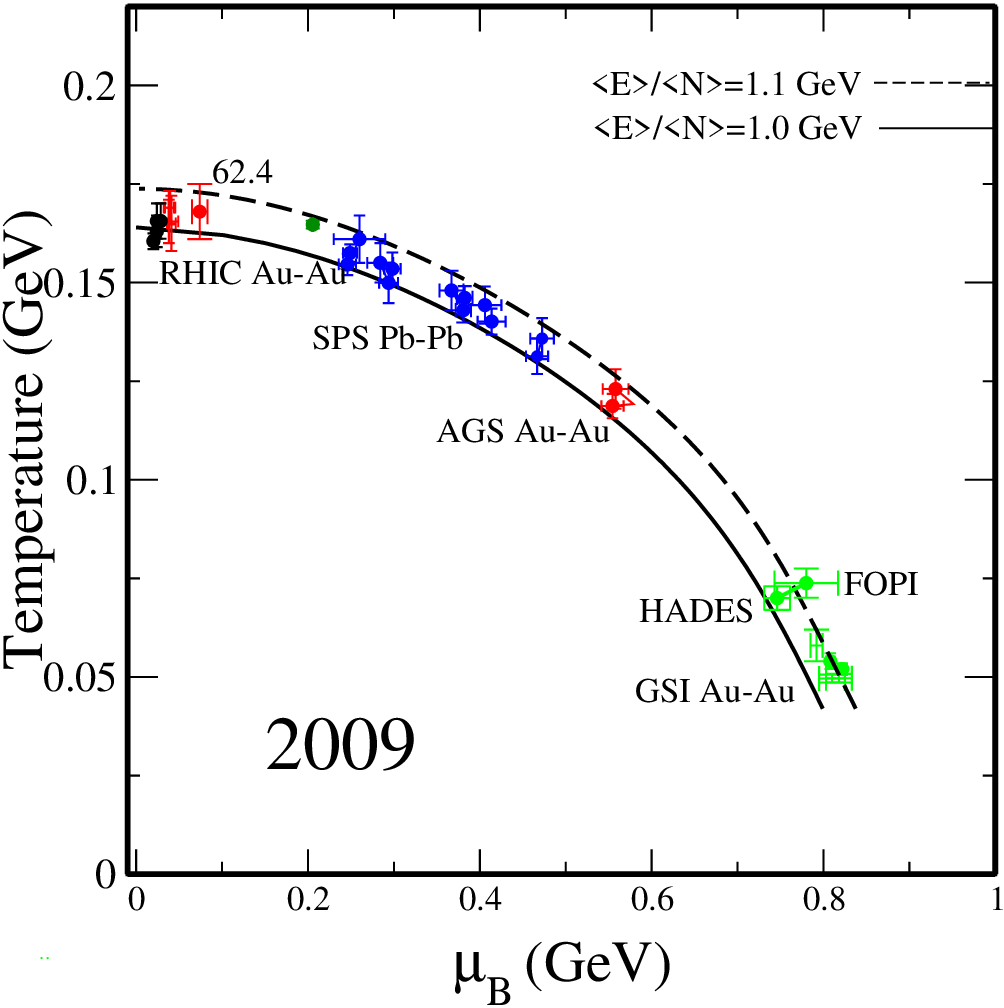,width=7.5cm}}
\caption{Results for the chemical freeze-out
temperature and baryon chemical potential. Curves obtained for constant
values of $E/N$ = 1.0 (full line) and 1.1 GeV (dashed line) are also
shown~\cite{1gev}.}
\label{eovern}
\end{figure}
%
%
A summary of the Statistical Model results  for the chemical freeze-out
temperatures and baryon chemical potentials 
at various energies is presented
in Fig.~\ref{eovern}. The freeze-out line shows a smooth change in going
from the low energies at SIS to the highest available energies at
RHIC, some of the more recent entries were obtained
 in~\cite{takahashi,schmah,gasik}.
Fig.~\ref{muB_T_e} shows these values as a
function of $\sqrt{s_{\rm NN}}$ exhibiting for \T~ a rising curve
which saturates above top SPS energies at a value of about
170 MeV. Note, that  the parameters $T$ and $\mu_B$  show a smooth change
with incident beam energy.
The values obtained can be parameterized as
\begin{equation}
T(\mu_B) = a - b\mu_B^2 -c \mu_B^4 . \label{Eqn:T(muB)}
\end{equation}
where $ a =  0.166 \pm 0.002$ GeV, $b = 0.139 \pm 0.016$
GeV$^{-1}$ and $c = 0.053 \pm 0.021$ GeV$^{-3}$,
and
\begin{equation}
\mu_B(\sqrt{s}) = \frac{d}{1 + e\sqrt{s}}, \label{Eqn:MuB(s)}
\end{equation}
with $d = 1.308\pm 0.028$ GeV and  $e = 0.273 \pm 0.008$
GeV$^{-1}$~\cite{Cle:2006}.
\begin{figure}[ht]
\begin{center}
\includegraphics*[width=7.5cm]{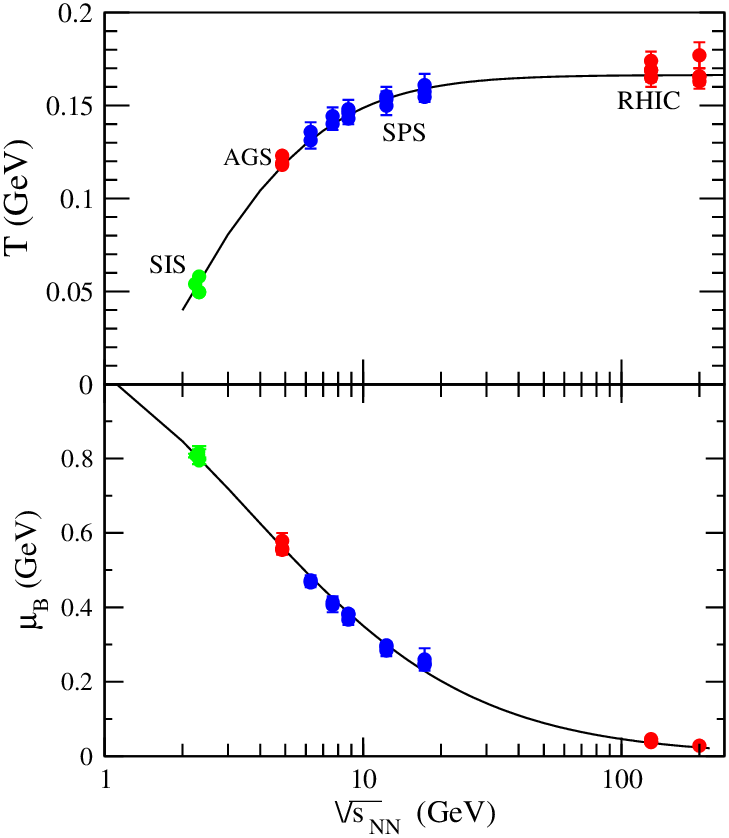}
\end{center}
\caption{Energy dependence of the chemical freeze-out parameters
$T$ and $\mu_B$. The curves have been obtained using a
parameterization discussed in the text. }
\label{muB_T_e}
\end{figure}
\section{Maximum relative strangeness content. }
\begin{figure}
\resizebox{0.985\textwidth}{!}{%
\includegraphics{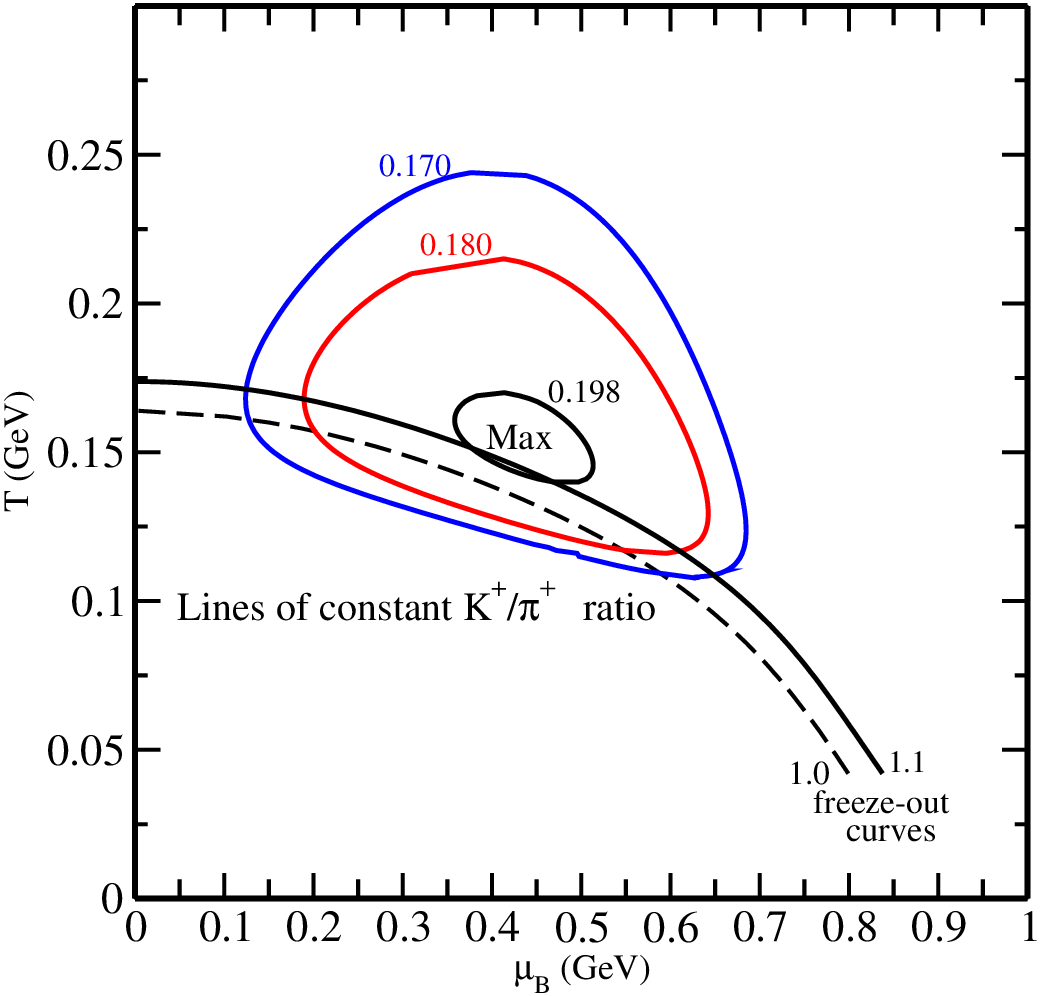}\hspace{5mm}
\includegraphics{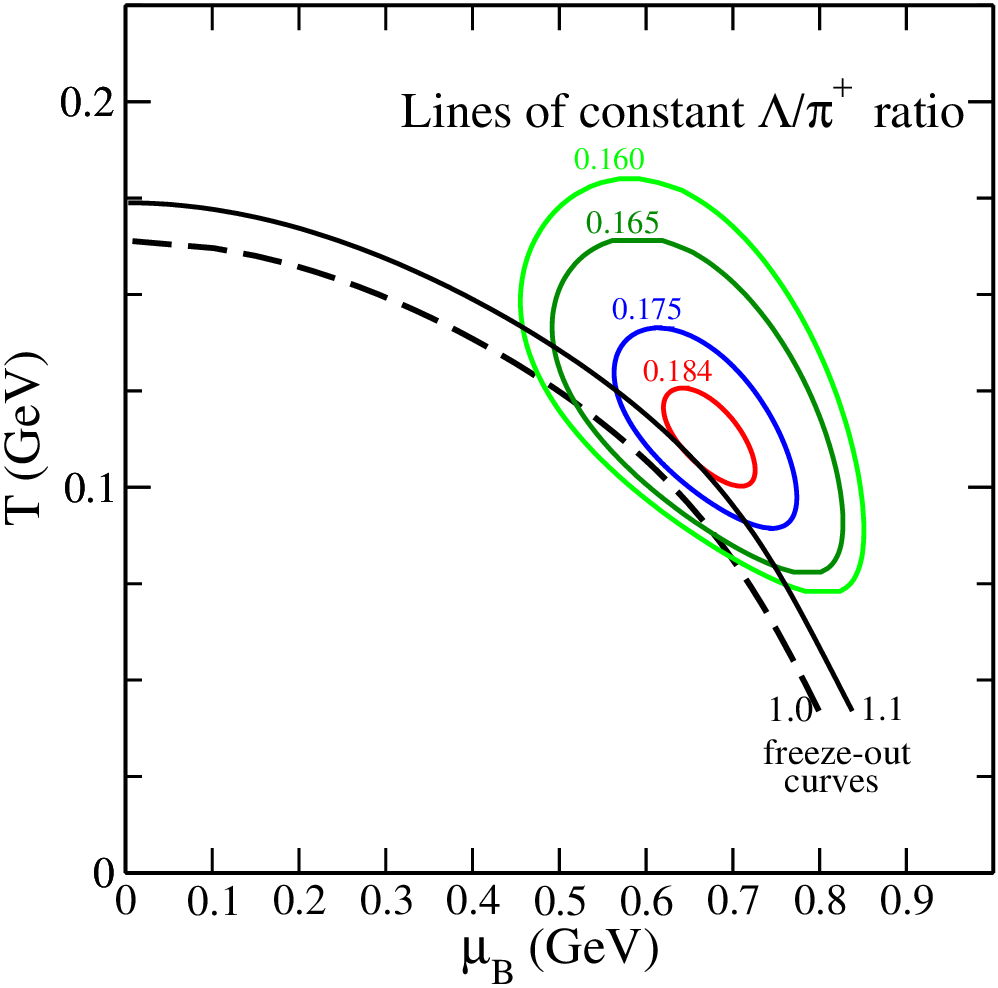}}
\caption{Contours of constant values of the  $K^+/\pi^+$ ratio (the left-hand figure) and the 
$\Lambda/\pi$ (the right-hand figure) ratios in the $T - \mu_B$ plane.
Calculations have been done using {\sc Thermus}~\cite{thermus}}\label{kplus_maxima}
\end{figure}

It is important to note that only the $K^+/\pi^+$ ratio exhibits
a sharp maximum while the $K^-/\pi^-$ ratio shows a continuous
rise with $\sqrt{s_{\rm NN}}$.
Contours of constant values of the $K^+/\pi^+$
and $\Lambda/\pi$  ratios  in the $T - \mu_B$ plane as calculated in the Statistical Model are shown in
Fig.~\ref{kplus_maxima}. The  $K^+/\pi^+$ ratio
exhibits a clear maximum just beyond the chemical freeze-out curve. Note that the position 
of the maximum in the $\Lambda/\pi$  ratio
is different from   that obtained for the 
 $K^+/\pi^+$ ratio. 
Again the maximum of the $\Lambda/\pi$  ratio is just beyond the line of
chemical freeze-out.
%
%
%
%
%

The Statistical Model describes the observed trends
qualitatively, but not the sharp maximum in $K^+/\pi^+$. Recently,
the model has been extended to include higher
resonances~\cite{Andronic:2008gu}. As these mostly decay into pions
the strong drop of the $K^+/\pi^+$ ratio towards RHIC energies as
observed in the data, is now better described.
The appearance of the maxima can be traced  to the specific
dependence of $\mu_B$ and $T$ on the beam energy as also pointed
out in Ref.~\cite{SK}.
The production of strange baryons
dominates at low $\sqrt{s_{\rm NN}}$ and loses importance at high
incident energies when the yield of strange mesons increases.
However, strange mesons also exhibit a maximum, albeit less
pronounced. This is due to the fact that strangeness production at
the lower energies occurs via the associated production,
i.e.~$K^+$ are created together with
hyperons~\cite{Cleymans:2004bf}. Therefore the $K^+$ mesons are
affected by the properties of the baryons, but the $K^-$ are not.
\begin{figure}[h]
\begin{center}
\centerline{\epsfig{file=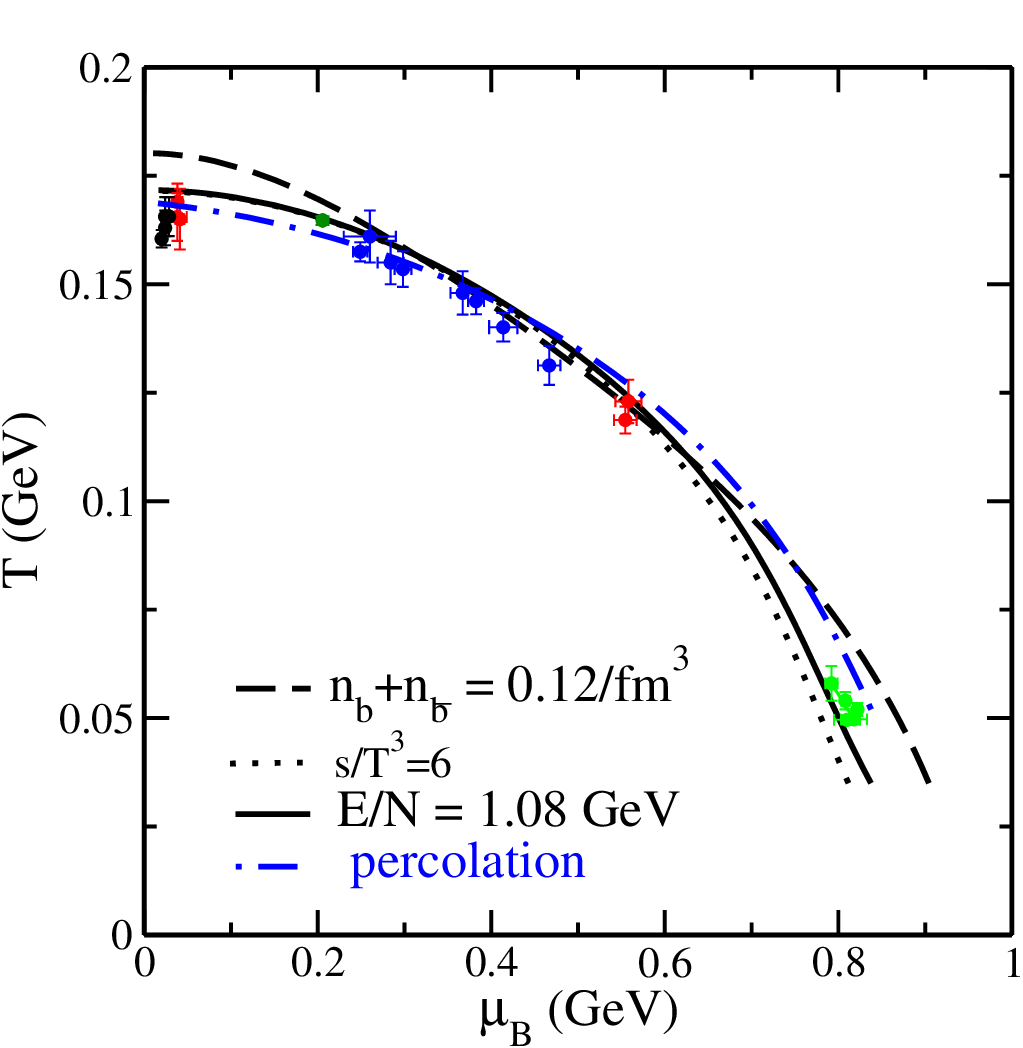,width=7.5cm}}
\caption{Comparison of model descriptions~\cite{Cle:2006} of the
chemical freeze-out  with results obtained from fits to the data using the
Statistical Model.}
\end{center}
\label{f11}
\end{figure}
%
%
\section{Transition from Baryonic to mesonic freeze out}
\begin{figure}
\resizebox{0.985\textwidth}{!}{%
\includegraphics{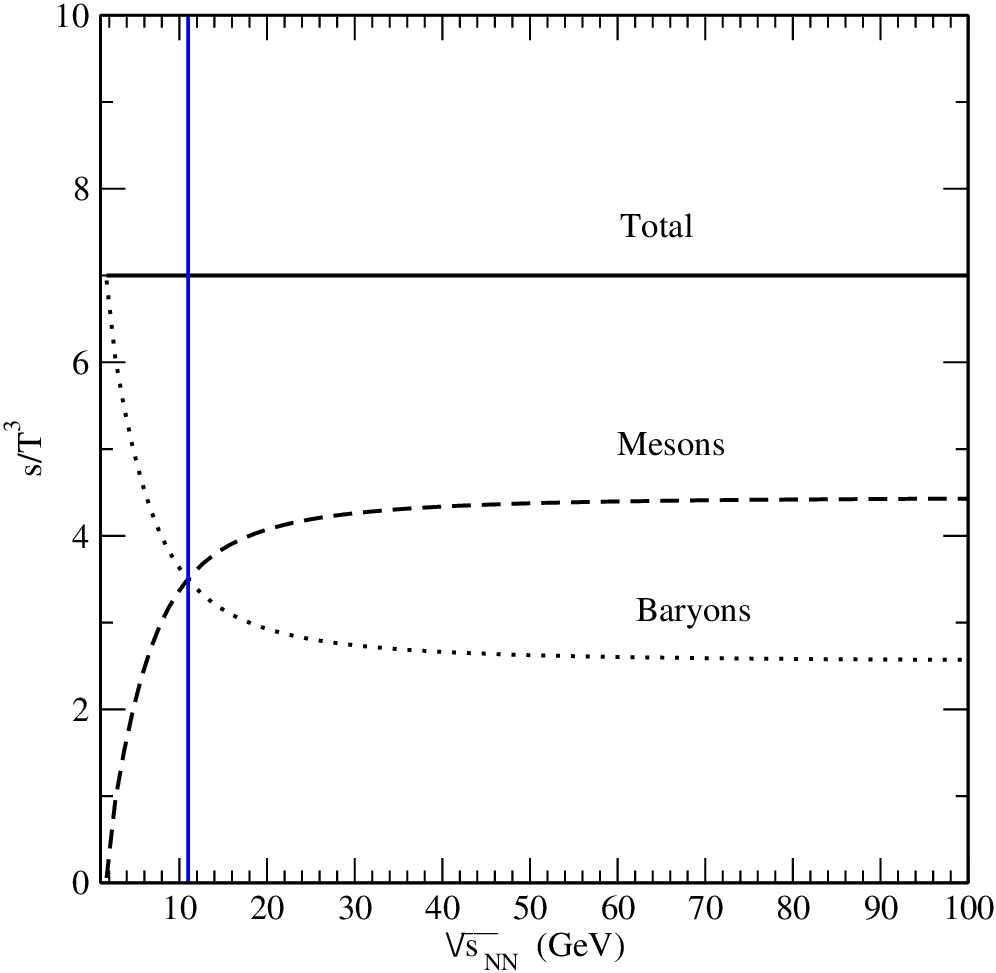}\hspace{5mm}
\includegraphics{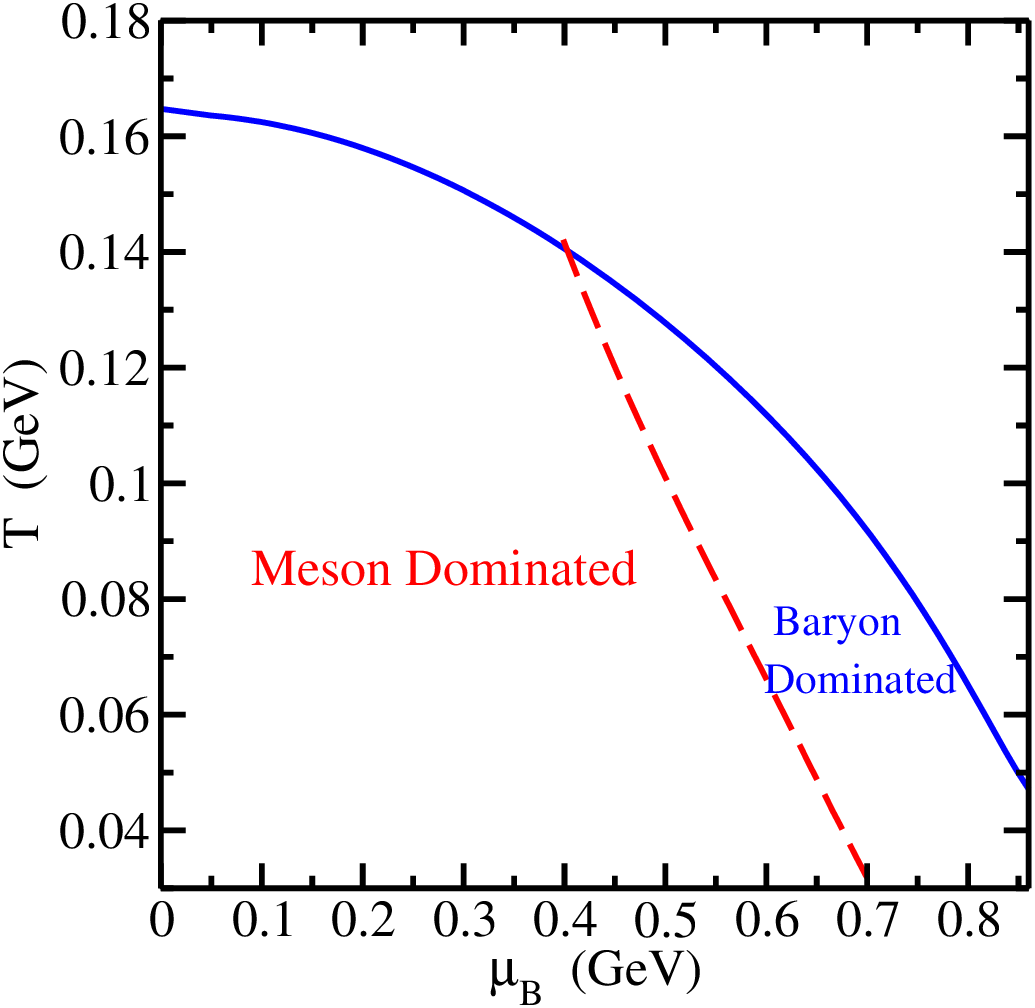}}
\caption{The left hand figure: the entropy density normalized to $T^3$
as a function of the beam energy.  The right hand figure: the line separating the $T-\mu_B$ plane into an area dominated by baryonic and one by mesonic freeze out.
The calculations within the Statistical Model have been done using {\sc Thermus}~\cite{thermus}.}\label{sovert3}
\end{figure}

While the Statistical Model cannot fully explain the sharpness of
the peak in the $K^+/\pi^+$ ratio, there are nevertheless several
phenomena giving rise to the rapid change which warrant a closer
look at the model.
%
%
%
%
%
As shown in Fig. 4  a constant value for the ratio  $s/T^3$ = 7
is a fairly good criterium to
describe the freeze-out curve~\cite{Cle:2006}.
We use it here
to describe the nature of the rapid change in the various ratios.
Fig.~\ref{sovert3}-left  shows  the entropy density divided by $T^3$ as a
function of beam energy as solid line.
The separate contributions of mesons and of baryons to the total
entropy are also shown by the dashed and dotted
lines. There is a clear change of baryon to meson dominance, using
the latest input of the Particle Data Group, including the
heavy resonances, this is estimated to happen around
$\sqrt{s_{\rm NN}}$ = 11 GeV.
Above this value the entropy is
carried mainly by mesonic degrees of freedom. It is remarkable
that the entropy density divided by $T^3$ is almost constant for
all incident energies  above the top AGS.

The separation line between meson dominated and baryon dominated
areas in the $T-\mu_B$ plane is given in
Fig.~5-right~\cite{transition}.
In this figure
the separation line crosses the freeze-out line at the stated
$\sqrt{s_{\rm NN}}$. This figure invites further speculations
as e.g.~an existence of a triple point.
%
%

\section{Summary}
It has been shown~\cite{max_strange} that the Statistical Model yields a
maximum in
the relative strangeness content around 30 $A$ GeV. This is due to
the saturation of the temperature $T$ while the chemical potential
keeps decreasing with incident energy. Since the chemical
potential plays a key role, it is clear that baryons are strongly
affected. Indeed, all hyperon/$\pi$ ratios yield maxima. In
contrast, the $K^-/\pi^-$ ratio shows a continuously rising curve
as expected. The $K^+/\pi^+$ ratio,
however, exhibits a maximum 
at the lower incident energies as  hyperon/$\pi$.  The model 
predicts that for different hyperon/$\pi$ ratios the maxima occur
at different energies. If experiments prove this, the case for a
phase transition is  weak.

The energy regime around 30 $A$ GeV seems to have specific
properties. It has been  shown that the entropy production occurs below
this energy mainly via creation of baryons, while at the higher
incident energies meson production dominates.

This work was supported by the German Ministerium f\"ur Bildung
und Forschung (BMBF), the Polish Ministry of Science (MEN) and by  the Alexander von Humboldt
Foundation (AvH).

\vskip 1.0cm

\end{document}